\documentclass[fleqn,10pt]{wlscirep}
\usepackage[utf8]{inputenc}
\usepackage[T1]{fontenc}

\usepackage{subfig}
\usepackage{xcolor} 
\title{Physics-assisted machine learning for THz spectroscopy: sensing moisture on plant leaves}

\author[1]{Milan Koumans}
\author[1]{Daan Meulendijks}
\author[1]{Haiko Middeljans}
\author[1]{Djero Peeters}
\author[2]{Jacob C. Douma}
\author[1,*]{Dook van Mechelen}
\affil{Eindhoven University of Technology, Department of Electrical Engineering, Eindhoven Henrik Casimir Institute, 5600 MB Eindhoven, the Netherlands}
\affil[2]{Centre for Crop System Analysis, Wageningen University, 6700 AK Wageningen, the Netherlands}
\affil[*]{\small \textcolor{blue}{correspondence to: j.l.m.v.mechelen@tue.nl}}

\begin{abstract}
Signal processing techniques are of vital importance to bring THz spectroscopy to a maturity level to reach practical applications. 
In this work, we illustrate the use of machine learning techniques for THz time-domain spectroscopy assisted by domain knowledge based on light-matter interactions. We aim at the potential agriculture application to determine the amount of free water on plant leaves, so-called leaf wetness. This quantity is important for understanding and predicting plant diseases that need leaf wetness for disease development. The overall transmission of a moist plant leaf for 12,000 distinct water patterns was experimentally acquired using THz time-domain spectroscopy. We report on key insights of applying decision trees and convolutional neural networks to the data using physics-motivated choices. Eventually, we discuss the generalizability of these models to determine leaf wetness after testing them on cases with increasing deviations from the training set. 
\end{abstract}
\begin{document}

\flushbottom
\maketitle

\thispagestyle{empty}

\section*{Introduction}
The quest to implement societally and industrially relevant applications of THz technology is impeded by aspects such as costs and performance as compared to alternatives. Despite the wide range of exploratory studies of THz technology since the late 1980s, and the numerous suggested applications,\cite{koch_terahertz_2023} very few products exist that use THz-based techniques. Often demonstrators underperform compared to a cheaper alternative that already exists or that is conveniently adapted to a new application. However, since THz technology is rather new on the market, costs will remain high at least for a while. The only promising way out is to find a novel application for the technology, which it can almost uniquely serve, and for which the business case is strong enough to support the high costs.\cite{van_mechelen_industrial_2015} In this case, its overall performance needs to justify its usage. On the hardware side, technological maturity has strongly improved over the last decades, although the progress at an integrated level is lagging.\cite{leitenstorfer_2023_2023} On the software side, despite crucial advancements such as model-based analysis that is now widely employed,\cite{van_mechelen_stratified_2014} the limitations are generality, robustness, and speed, which are essential for realistic application cases. 

The agriculture sector is an area where a multitude of sensing technologies are employed to aid management decisions. Certainly due to this reason, there is a vivid interest in searching for applications in this field where THz spectroscopy can make a difference.\cite{s_recent_2022} Globally, pests and pathogens are a big threat to crop production, with yield losses reported in the range of $9-21$\;\%.\cite{oerke_crop_2006}  For some pathogens, such as water molds and some fungi, the presence or absence of free water on the surface of leaves, so-called leaf wetness, is key for infection and/or sporulation and is therefore an important parameter in disease epidemiology. A famous example is \textit{Phytophthora infestans}, the causal agent of potato late blight, responsible for the Irish Potato Famine in the mid-19th century. The development of \textit{P. infestans} depends on the presence of leaf wetness and the surrounding temperature.\cite{bregaglio_multi_2011,huber_modeling_1992} In ideal circumstances, \textit{P. infestans} can decimate a potato crop in less than 10 days. Control of late blight, as well as that of other pests and pathogens, is nowadays mostly done using crop protection products.\cite{goffart_potato_2022}  There is, however, a strong push from policymakers to reduce this. Early detection and improved predictions of when and where diseases may be expected can help targeted (preventive) measures. Instead of directly detecting the molds, which in the field are difficult to observe, predicting leaf wetness during the growth season is an important input for decision support systems to advise on spraying crop protection products.

Terahertz spectroscopy is particularly suited to accurately sense little amounts of water and is proven to probe leaf properties.\cite{li_non-invasive_2020,gente_determination_2013,gente_monitoring_2015,singh_three-dimensional_2020} Other technologies that can sense water, such as visible imaging, will have difficulties estimating very small amounts of water due to a lack of contrast. Currently used electric measurements are very local, not directly performed on a plant leaf, and lack interpretability.\cite{rowlandson_reconsidering_2015} Although several strong water absorptions are situated in the infrared spectral range, its frequency domain operation will make it difficult to distinguish between surface and content water of a leaf, and its shorter wavelength also causes more local sensing.

To accurately extract parameters from a canopy of plant leaves, which may move, are curved, have an anisotropic structure, and are in an environment that is determined by weather conditions, a suitable signal processing method is needed. Model-based signal processing, e.g., based on the transfer matrix method, can be powerful in well-behaved conditions, such as paint layer inspection in an automotive paint shop and wafer metrology in a cleanroom,\cite{van_mechelen_thickness_2021} it will not describe well the complex situation at hand. Signal processing using a data-based learning method could be the solution in this case. Many other studies have used machine learning on THz spectroscopic data.\cite{park_machine_2021,wang_terahertz_2020,cao_terahertz_2020,wang_automatic_2021} However, most of these studies suffer from a lack of transparency of the used method such that the quality of the result is unclear in aspects of generality and reproducibility.

Here, we study the application of machine learning models to THz spectroscopic data of moist leaves to predict leaf wetness. The primary focus of this paper is to present a clear and concise approach for applying decision trees and convolutional neural networks to THz data. Based on the light-matter interaction, we motivate the architecture of the used models and give particular emphasis to feature engineering. Both methods accurately predict the amount of water on a leaf, independent of the droplet pattern. Eventually, we discuss the influence of model variability and deduce their generalizability.

\begin{figure}%
    \centering
    {{\includegraphics[width=.9\textwidth]{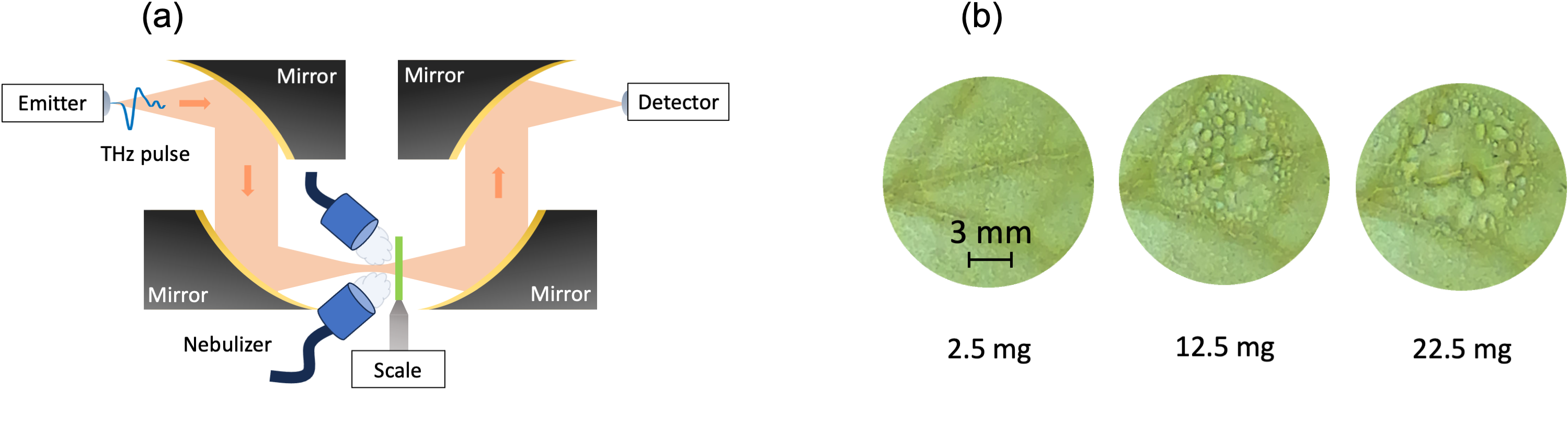} }}%
    \qquad
    {\includegraphics[width=.9\textwidth]{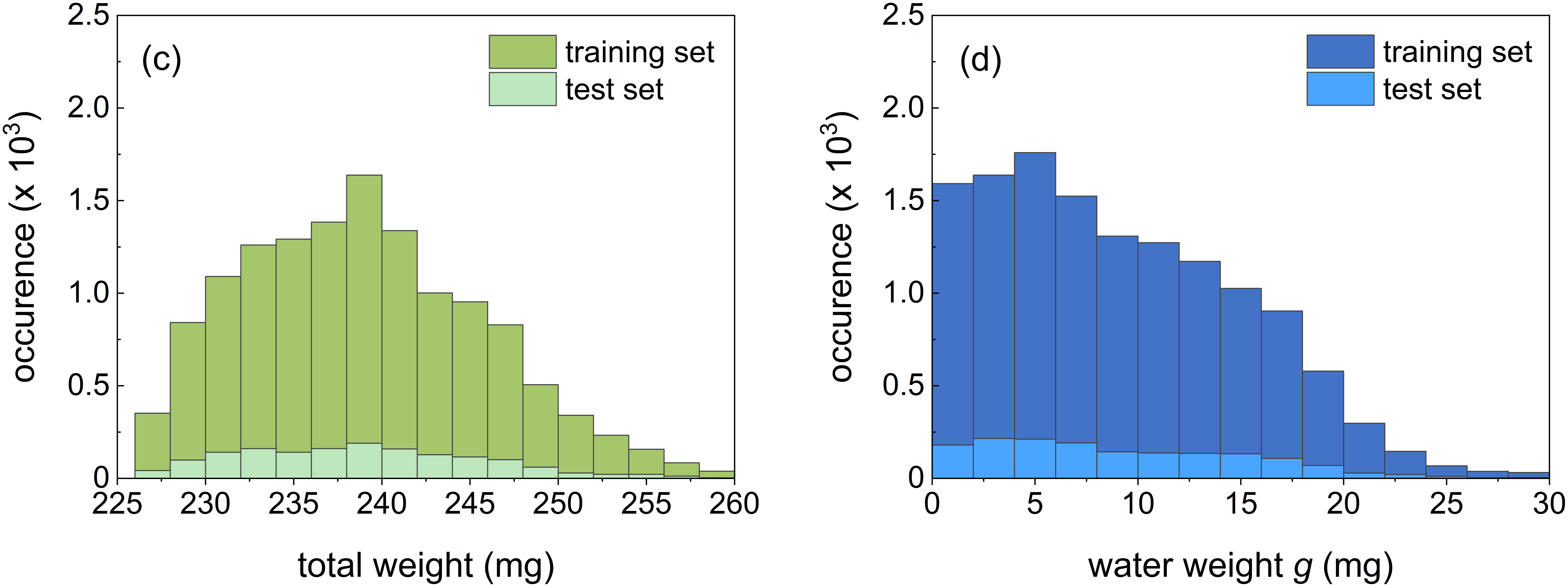}}%
    \caption{(a) Experimental setup showing a leaf on a scale in a THz time-domain transmission configuration. (b) Selected images from a measurement series showing the formation of a droplet pattern with weight $g=2.5$, $12.5$ and $22.5$ mg, respectively. (c)\,Distribution of the experimental data $\mathbf{E}(t)$ categorized by the total gravimetric weight comprised of the plasticized leaf and water pattern weight $g$. (d)\,Distribution of the same data as in panel (a) but here categorized by $g$.}%
    \label{fig:image01}%
\end{figure}

\section*{Results}
\begin{figure}
    \centering
  	\includegraphics[width=.9\textwidth]{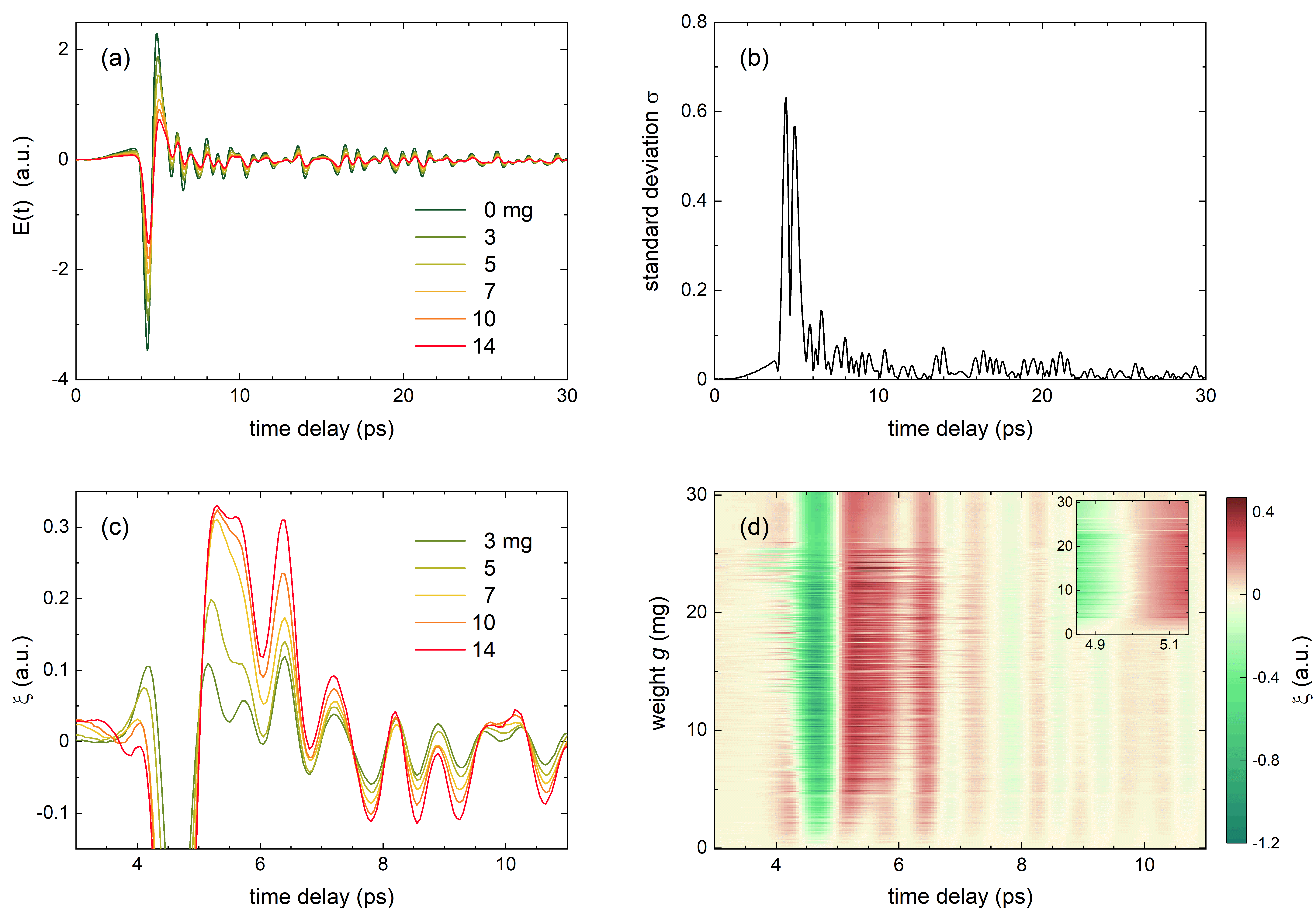}
    \caption{(a) Experimentally determined $E(t)$ for selected values of $g$ in ambient conditions (see methods). (b) Standard deviation $\sigma(t)$ of $\mathbf{E}(t)$. (c)~$\xi=E(t)/\max(E(t))-E_0(t)/\max(E_0(t))$ with $E_0(t)$ the transmission for $g=0$. This quantity provides detailed insight into temporal ranges of large variation. (d)~$\mathbf{\xi}$ vs. $g$. The inset highlights the temporal shift around 5\,ps as a function of $g$. }
    \label{fig:image02}
\end{figure}

The transmitted electric field $E(t)$ of water patterns on plasticized plant leaves has been experimentally recorded in transmission geometry at THz frequencies ($0.1-3$ THz) in the time-domain for about 12,000 distinct patterns as described in the methods (Fig.~\ref{fig:image01}a,b). We recorded two data sets, $\mathbf{E}(t)$ for which the water patterns are deposited on the top side of the leaf, and $\mathbf{E'}(t)$ for which the water patterns are on the bottom side of the leaf. For $\mathbf{E}(t)$, the distribution of the total sample weight, composed of the leaf and the water pattern on top, manifests a roughly normal distribution (Fig.~\ref{fig:image01}c), whereas the distribution of only the water weight $g$ is predominant for low weights (Fig. \ref{fig:image01}d). The maximum value of $g$ is given by the condition where droplets run down the leaf. Fig.~\ref{fig:image02}a shows a given measurement series $E(t)\in\mathbf{E}(t)$ for selected values of $g$, which visually drastically changes with $g$. Besides, each $E(t)$ is also related to a given water pattern with its droplet size distribution. Therefore, spectra $E(t)$ with identical value $g$ may mutually differ.  The standard deviation $\sigma(t)$ of $\mathbf{E}(t)$ demonstrates that the influence of water on $E(t)$ is predominantly present in the range (4, 5) ps, related to a single passage of THz radiation through the droplet pattern and leaf (Fig. \ref{fig:image02}b). However, maybe of more importance for sensing water is the range of $E(t)$ where the radiation internally reflects within the water droplets. From simulating $E(t)$ using a transfer matrix model, we conclude that the first internal reflection inside droplets shows up in $E(t)$ as a shoulder right after the largest positive peak, that is, around 5.5\,ps. This can be better visualized through the quantity $\xi=E_i(t)/\max(E(t))-E_0(t)/\max(E_0(t))$ with $i$ the number of acquisitions within a measurement series, and where $i=0$ stands for a leaf with $g=0$ (see Fig. \ref{fig:image02}c). When plotting $\xi$ for the entire data set $\mathbf{E}(t)$ as a function of $g$ (Fig. \ref{fig:image02}d), this effect shows up a fork for low $g$ around 5.5\,ps. The ray trajectory in the droplet-plastic-leaf-plastic system is, however, sufficiently complex that more subtle features in $E(t)$ cannot be uniquely attributed to a specific radiation path. To obtain an accurate prediction of the leaf wetness $g$, we employ two different data-driven methods, decision tree regression and convolutional neural networks. For each algorithm, we determine the mean absolute error and median percentage difference, defined as the median of $|g_p-g_b|/(g_b + \epsilon)$ with predicted weight value $g_p$, benchmark weight value $g_b$, and small $\epsilon$ for stability when $g_b\approx0$.

\subsection*{Decision Trees}
The above-described problem to predict $g_p$ from the experimental input matrix $\mathbf{E}(t)$ is equivalent to determining a function $f$ for which holds that $f: \mathbf{E}(t) \rightarrow g$. We have ascertained that the relation between distinct features of $\mathbf{E}(t)$, which will be discussed below, and the target variable $g$ is not linear. To evaluate this problem, we choose a decision tree method which is proven to work with non-linear mappings $f$ for mid-sized data sets such as that of $\mathbf{E}(t)$. 

\begin{figure}%
    \centering
    {{\includegraphics[width=.9\textwidth]{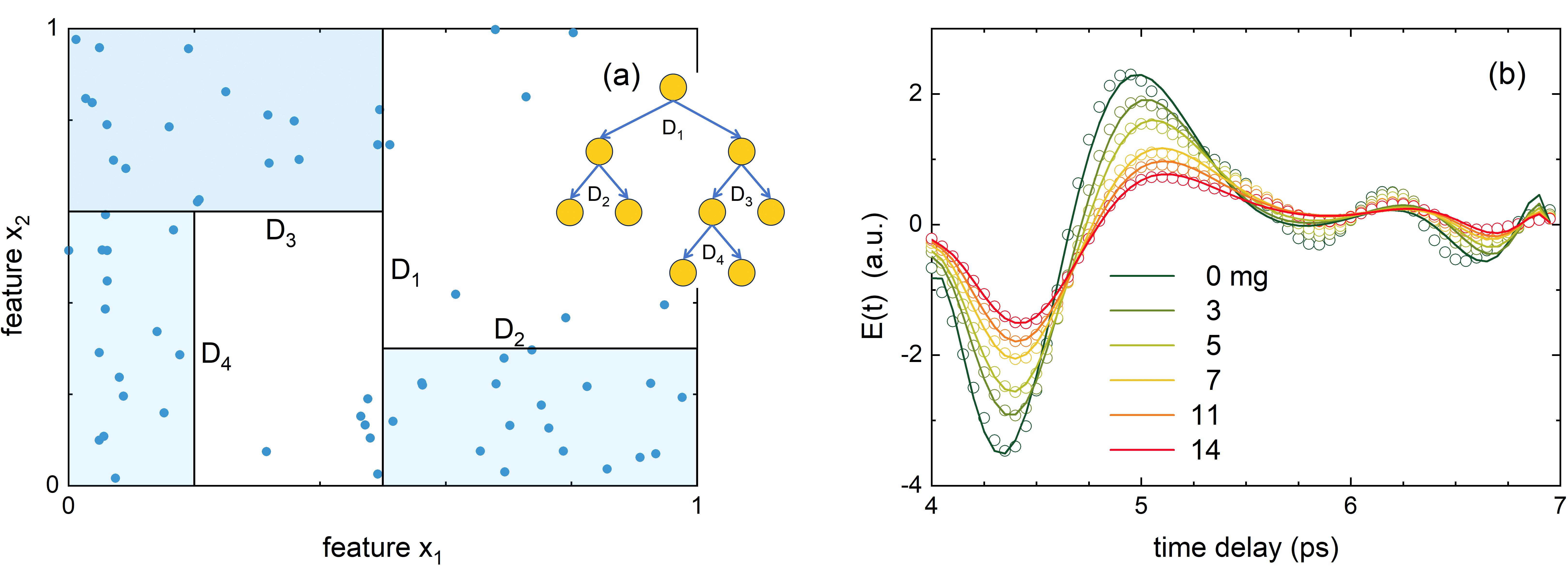}}}%
    \caption{(a) Schematic view of the mechanism of a decision tree. The dots represent data containing features $x_1$ and $x_2$, and the black lines show the division of the parameter space according to split criteria D$_s$ with $s=1..4$. The predicted value of each section is the average of the benchmark values in the corresponding division. (b) Experimentally determined $E(t)$ for selected values of $g$ (circles) together with polynomial fits with $n=11$ (solid lines).}%
    \label{fig:image03}%
\end{figure}

A decision tree is a supervised machine learning method that is shaped as a flowchart in which iterative decisions lead to a piece-wise approximation of the target variable (see Fig. \ref{fig:image03}a for a single class and two features $x_1$ and $x_2$).  Decisions within regression trees use quantitative split criteria such as the absolute error (L1-norm) or squared error (L2-norm) to determine the best split at a decision point D. A single decision tree is deterministic in its prediction, making it prone to overfitting. To overcome this shortcoming, an ensemble of simplified trees is used. Each simplified tree independently predicts the target value, after which the predictions are averaged. This reduces the variance of the model while maintaining the predictive power of the decision trees. Simplifying the trees by limiting the seen input data is called bagging and is a well-proven method that is capable of learning nonlinear relations between input features and the target variable. Due to the simplistic nature of decision trees, inference is interpretative and most importantly, ensemble modeling leads to improved robustness compared to single estimators.
As common for machine learning algorithms, decision trees fit on features that are reminiscent of the data. Many reports employing machine learning methods on THz time-domain data either first convert to the frequency domain, thereby losing crucial spatial information, or simply use the time trace $E(t)$.\cite{li_nondestructive_2022,wang_terahertz_2020,cao_terahertz_2020} However, points within $E(t)$ are time-correlated and when directly used as features, the correlation may hamper determination of the independent effect of each feature on the target variable. We therefore use feature engineering to extract the most reminiscent features from $E(t)$, which are not temporally correlated. We hereto fit a polynomial function of degree $n$ to relevant temporal ranges of $E(t)$ (Fig.~\ref{fig:image03}b), as further detailed in the methods. $n$ is chosen as small as possible, but large enough to capture reminiscent features of $E(t)$ to best predict $g$. Although visually the fit may not seem optimal, for the algorithm it is. Besides the coefficients of the polynomial terms, also the beginning of the time window $t_{\text{start}}$ is added as a feature as well as the absolute air humidity $a$, which has a strong influence on $E(t)$. Optimization for the displayed range in Fig.~\ref{fig:image03}b leads to $n=11$ as described in the methods. The feature vector thus reads [$t^0$, $t^1$, \ldots, $t^{11}$, $t_{\text{start}}$, $a$], where $t^0$ is the bias term and will be further indicated like that. Among these features there can be, however, ones that are mutually correlated and ones that are only a little related to the target variable. We hereto employ recurrent feature elimination to reduce the dimensionality of the feature matrix and enhance the performance of the model, as further detailed in the methods. Eventually, hyperparameters responsible for the regularization of the final ensemble model are determined using 5-fold cross-validation on a training set consisting of 85\,\% of the total data set $\mathbf{E}(t)$.

The predictive performance of a bagged decision tree as described here and further detailed in the methods was evaluated on an unseen test set consisting of 15\,\% of the total data set $\mathbf{E}(t)$ and is shown in Fig.~\ref{fig:image05}a. The mean absolute error on $g_p$ is $0.35^{+0.17}_{-0.1}$ mg and the median percentage difference is $3.4^{+1.5}_{-1} \%$. The mean inference time is 139 ms ($\sigma = 22$ ms) per sample using the hardware as mentioned in the methods. The indicated error bars are motivated in the Discussion.

\begin{figure}%
      \centering
    {{\includegraphics[width=\textwidth]{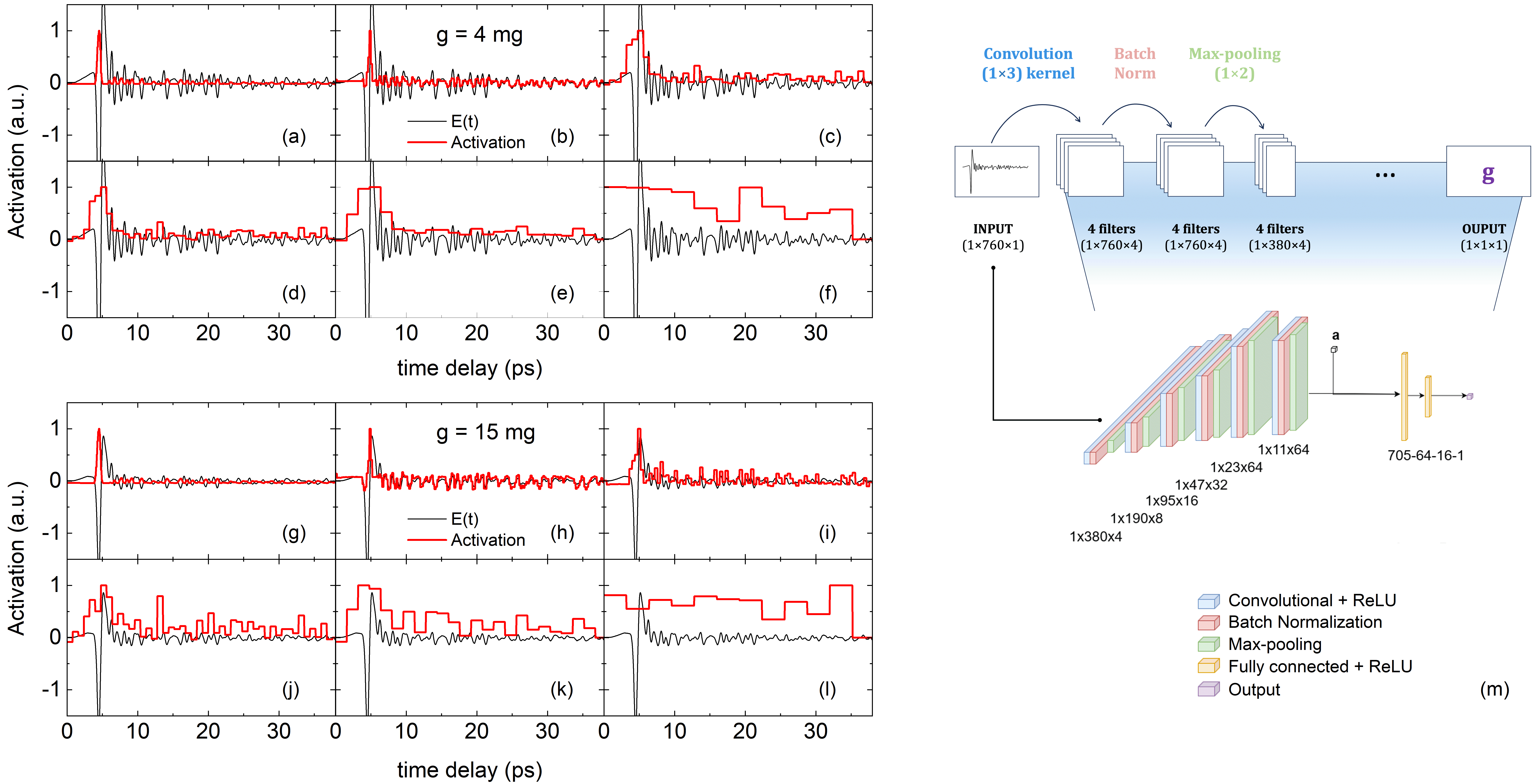}}}%
    \caption{Activation of $E(t)$ per convolutional layer $\ell$ for (a-f) $g=4$\,mg and (g-l) $g=15$\,mg averaged over all feature maps (panels (a,g) correspond to $\ell=1$ etc.). (m) Architecture of the convolutional neural network, indicating the output dimensions for each layer. The dimensions after each max-pooling layer are given as $1\times d_\ell\times k_\ell$ where $d_\ell$ is the length of layer $\ell$ and $k_\ell$ the number of filters of layer $\ell$.}
    \label{fig:image04}%
\end{figure}

\subsection*{Convolutional Neural Network}
An artificial neural network is a convenient tool to autonomously discover intricate patterns and representations of signals. The simplest neural network predicts an output variable $y_i$ given input data $X_{train}$ and known target variable $y_{train}$ as in linear regression by learning a mapping function $f:X_{train} \rightarrow y_{train}$. The architecture of such a neural network consists of an input layer with input data $x_i \in X_{train}$ and an output layer with output data $y_i$. The real strength of a neural network is, however, to find patterns and representations in high-dimensional input data. Hereto, cross-correlations between the inputs are learned using so-called perceptrons (artificial neurons) that output a latent variable $h$ depending on input $x_i$. Subsequently, an activation function, for instance $\varphi(h)=\max(0,h)$, is applied to $h$. This has the effect that only neuron output of sufficient importance is fed deeper into the network and correlated to the output of neurons in the subsequent layer. In this way, the network can learn non-linear relationships beyond the simple perceptron model. In case the input can be represented as an image, patterns are more effectively learned by using a so-called convolutional neural network (CNN). In such a network a small kernel matrix is scanned over the image matrix to learn local relations between data points. For pictures, these kernel matrices can represent lines and circles, but also more complicated patterns, which are learned by the network. In addition, the method also works for correlated 1-dimensional signals.

The aim here is to construct a CNN, train it on the experimentally acquired time-domain data $\mathbf{E}(t)$, and compare its performance to predict $g$ to prediction using decision trees as discussed in the previous section. For the few studies where THz time-domain data is used for CNNs, rarely the full functional $E(t)$ is used despite its spectral richness and possibility to effectively augment the number of samples as mentioned in the Discussion.\cite{mao_convolutional_2020, wang_convolutional_2021,wang_automatic_2021} A typical CNN consists of many layers which can be grouped into a feature extraction part and a regression part. The feature extraction part consists of multiple convolutional layers $\ell$ that utilize kernel operations to convolve over the input vector, which enables the detection of pertinent features in signal $E(t)$. Early layers rather reveal the local context of signal $E(t)$, whereas deeper layers combine activations of different temporal regions to extract the global context of signal $E(t)$. The patterns of the kernels are determined through backpropagation as described in the methods. For each layer, the number of distinct kernels $k$, also called filters, has been empirically chosen as a function of the size of $E(t)$, the complexity of detected patterns in $E(t)$, and the condition to keep the network size as small as possible. Hence, to capture basic patterns, layer $\ell=1$ of our CNN has four filters, each of dimension $(1\times3)$ applied to each $E(t)$ of dimension $(1\times760)$ (see Fig.~\ref{fig:image04}m). This results in a feature map of dimension $(1\times760\times4)$. Fig.~\ref{fig:image04}a,g displays the activation of $E(t)$ for $g=4$ and 15\,mg, respectively, that is, the average of the four filters projected onto $t$. Similar to model-based signal processing of $E(t)$,\cite{koumans_sensing_2022} CNN shows the largest activation in the range around the absolute minimum of $E(t)$ and hence demonstrates the local character of this first layer. Batch normalization has been applied after each convolutional layer, which normalizes the features during forward propagation. This ensures activation throughout a deep network and results in improved convergence while it simultaneously works as a regularizer.\cite{ioffe_batch_2015} A so-called max-pooling layer is inserted behind each normalization layer to reduce the network size by downsampling the resulting feature maps. Conceptually, this layer structure increases the receptive field of each neuron. $\ell$ should be chosen such that the receptive field of neurons in the last layer covers the complete input signal $E(t)$. In our case, we empirically evaluated that $\ell=6$. We can verify the feature extraction performance of this architecture by inspecting the activations for two cases of $g=4$ and 15\,mg (see Fig.~\ref{fig:image04}a-l). With increasing $\ell$, the activation loses its local character and spreads throughout the entire temporal range. The concomitant widening and increasing height of the block functions is the effect of the max-pooling operations, where every iteration halves the time window and eliminates the smallest values. The activation shows that for earlier layers, the  range (4, 5)\,ps is of main importance for the network. This is in agreement  with the earlier observation that the largest amplitude of $E(t)$ varies most with $g$ (cf. Fig.~\ref{fig:image02}a), and as such is the most basic pattern of $E(t)$. Interestingly, for $\ell=3,4$ and 5, the activations show increased values also in the range (5,7) ps. As mentioned before, for $g<7$\,mg first internal reflections within the droplets occur around 5.5\,ps. In addition, from modeling $E(t)$ using the transfer matrix method, we find that the first internal reflection within the leaf material, although strongly damped, occurs around 7\,ps. It thus turns out that these regions which have an increased importance from a light-matter perspective are likewise important for a CNN. Moreover, the increased activity in other regions makes our CNN sensitive to details of $E(t)$ that can contain aspects that are difficult to incorporate into a physical model.

\begin{figure}
    \centering
    \includegraphics[width=.9\textwidth]{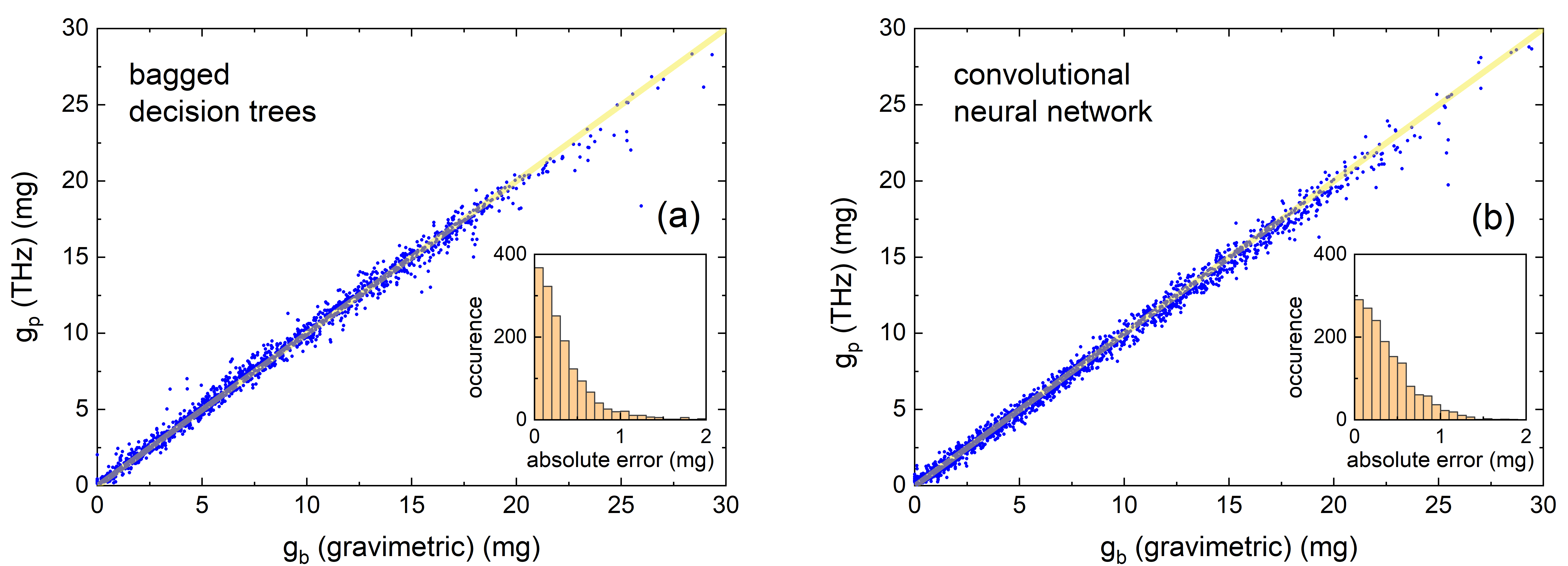}
    \caption{Predicted weight $g_p$ using (a) bagged decision trees and (b) a convolutional neural network, both versus the benchmark weight $g_b$. The insets display the distribution of the absolute error of $g_p$ for both models.}%
    \label{fig:image05}
\end{figure}

After feature extraction, regression is performed by a fully connected artificial neural network. Hereto, all previous activation layers are flattened to a 1D vector consisting of a concatenation of the 64\  $1\times11$ activations as well as the absolute humidity $\mathbf{a}$. The fully connected network consisting of two layers subsequently learns the nonlinear relation between determined input features and the target variable by minimizing its loss function given by the squared error $|\mathbf{g}_b - \mathbf{g}_p|^2$ using gradient descent optimization.

The CNN is trained on $\mathbf{E}(t)$ for 300 epochs with a batch size of 128 using a train-test ratio of 85:15, where an epoch is a single iteration for which the complete training set has been (forward and backward) propagated through the network. During training, a validation set of $10\%$ of the training set is used to validate the performance after each epoch. Fig.~\ref{fig:image05}b graphically shows the performance of the model, having a mean absolute error of $0.38\,^{+0.17}_{-0.1}$mg and a median percentage difference of $4.1^{+1.5}_{-1}\%$. The mean inference time is 70\,ms ($\sigma = 9$ ms) per sample using the hardware as mentioned in the methods.

\begin{figure}
    \centering
    \includegraphics[width=\textwidth]{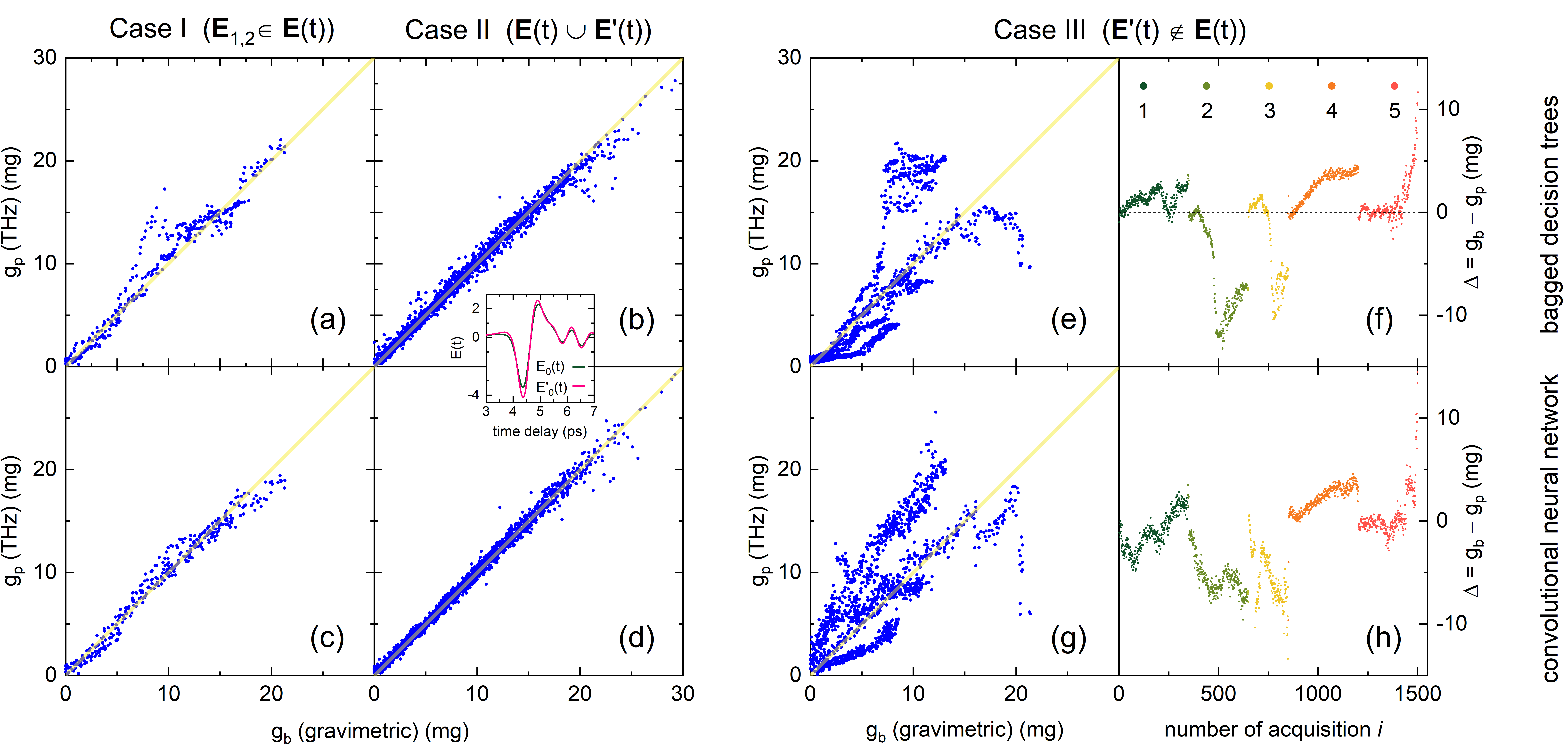}
    \caption{Predicted weight $g_p$ vs. benchmark weight $g_b$ using (a,b,e) bagged decision trees and (c,d,g) CNN, for three test cases. For each case, the test data set is mentioned in the figure title. Case I tests generalizability towards unseen water patterns using two unseen measurement series from $\mathbf{E}(t)$. Case II tests generalizability towards water patterns on top and below a leaf using the enlarged data set $\mathbf{E}(t)\cup\mathbf{E'}(t)$. Case III tests generalizability towards water patterns on an unseen leaf surface by training on $\mathbf{E}(t)$ and testing on $\mathbf{E'}(t)$). The inset displays $E(t)$ and $E'(t)$, both at $g=0$, indicating the difference in the transmitted electric field between irradiating a leaf from the top side and the bottom side, respectively. $\Delta=g_b-g_p$ for the 5 data sets (indicated by the color scheme) of $\mathbf{E}'(t)$ vs. the number of acquisitions $i$ for (f) bagged decision trees and (h) CNN. }%
    \label{fig:image06}
\end{figure}

\section*{Discussion}

The presented results demonstrate that using two very different data-driven approaches, leaf wetness can be determined using THz spectroscopy within a confidence interval of about 3 to 4\%, independent of the method. This is well beyond the performance of model-based analysis to determine leaf wetness using THz spectroscopy.\cite{koumans_sensing_2022} Upon comparison to methodologies reported for the closely related property of leaf water content,\cite{Jordens2009,gente_monitoring_2015,li_non-invasive_2020} we estimate that our study, with its large data set and comprehensible approach, gives a realistic view of the chances of performing quality control on plants.

Although most studies on machine learning conclude at this point, we are here interested in the generalizability of the models. Therefore, we consider the following further test cases. For case I, we use 37 out of the 39 measurement series of dataset $\mathbf{E}(t)$, train the learning methods according to the same procedure as described before, and test on the two remaining series. Both methods underperform as compared to the results shown in Fig.~\ref{fig:image05}, with decision trees having a median percentage difference of 8.8\,\% and CNN 6.7\,\% (see Fig.~\ref{fig:image06}(a,c)). The question thus poses about the origin of this decreased performance. The shape of $E(t)$ is both determined by $g$ and the water patterns, as mentioned before. However, the temporal range (4, 5)\,ps with the largest amplitudes is of leading importance for predicting $g$, and $\min_t \mathbf{E}(t)\approx5$\ ps manifests a monotonous variation with $g$ (see the inset of Fig.~\ref{fig:image02}d), independent of the water patterns, both suggesting that $g$ predominantly determines $E(t)$. Yet, the values of $g$ of the test set are a subset of the range of $g$ on which the models are trained. The effect of water patterns on $E(t)$, on the other hand, is more subtle and beyond the range of the largest amplitudes, as discussed before. We conjecture that the test set contains water patterns, rather than different values of $g$, that are unseen to both methods. The relative underperformance of decision trees is then probably directly linked to its mechanism for which a slight variation in $E(t)$, here due to a different pattern, alters the polynomial fits and thus features on which it has been trained. CNN on the other hand directly learns regions in $E(t)$ which are of relevance for learning $g$ for all trained water patterns. Case II considers $\mathbf{E}(t)$ as well as an additional 1501 time traces $E(t)$, referred to as $\mathbf{E'}(t)\notin\mathbf{E}(t)$ where the moisture pattern is created on the bottom side of the leaf. Instead of shining on the smooth and reflective top surface as for $\mathbf{E}(t)$, for $\mathbf{E'}(t)$ the beam now enters the leaf through the dull and rough bottom surface, thereby significantly changing $E(t)$ (see inset of Fig.~\ref{fig:image06}). Please note that the water pattern is always on the emitter side. The methods are trained on a subset of $\mathbf{E}(t)\cup\mathbf{E'}(t)$ and tested on an unseen subset of the same data set. Fig.~\ref{fig:image06}(b,d) shows a similar performance in predicting $g_p$ as when considering only $\mathbf{E}(t)$ (cf.~Fig.~\ref{fig:image05}), also indicated by the median percentage difference of 4.4\,\% for decision trees and 3.8\,\% for CNN. This indicates that the models can extract relevant features within $E(t)$ no matter the underlying leaf material. In case III, we train the models on $\mathbf{E}(t)$ and test them on the unseen data set $\mathbf{E'}(t)$. This would correspond to the practical situation where the method has learned based on water patterns on leaves which physically are not the same as those on which the model is inferred. The performance of the methods is rather lousy (Fig.~\ref{fig:image06}e,g), as indicated by the median percentage difference of 39\,\% for decision trees and 50\,\% for CNN. Fig.~\ref{fig:image06}f,h displays the absolute deviation $\Delta=g_b-g_p$ grouped according to the five measurement series of $\mathbf{E'}(t)$ as a function of the number of acquisition $i$. Each series starts at low $i$ with $g=0$ 
after which both $i$ and $g$ increase concomitantly. It turns out that $\Delta$ is small at low $g$ and deviates with increasing $g$. Signatures of water patterns with small droplets are thus overall well recognized by the models, despite the unseen patterns of $E(t)$ due to the flipped leaf. We conjecture that this sensitivity may be related to the presence of reminiscent features of $E(t)$ for $g<7$\,mg, due to internal reflections inside the droplets and leaf material (see Results), on which the network can train. These three test cases indicate that the learning model needs input training data that are quite close to the data set from which will be inferred. Although it is promising that $g$ is well predicted from data corresponding to water patterns that have not been seen before, unseen variations of the leaf material cause changes in $E(t)$ to which the methods are not robust. This could  be resolved by training the models on data that besides a large variation of $g$ and water patterns also include a large diversity of leaves.

The confidence and related error bars of each machine learning model also depend on specific parameters that are chosen within the architecture and implementation. For decision trees, for instance, we made use of domain knowledge to select four temporal regions within $E(t)$. However, a different choice of the temporal intervals will alter the optimal polynomial order $n$, modify the feature vector, and thus the performance of the method. For the convolutional neural network, the specific network architecture expressed by variables such as the number of layers, filters, and the number of epochs significantly varies the performance of the network. For the current data set, the quoted performance due to the described variations has an estimated error bar of the order of around $1.5\%$. However, also the data set size is an important parameter. Whereas for decision trees 9000 time traces with 11 features each gives a significant data set of $10^5$ features to train on, for CNNs the final feature size is more difficult to estimate. Although all temporal points of $\mathbf{E}(t)$ with size (9000, 760) are used as input, feature extraction leads to much fewer features than the number of elements of this matrix. By visual inspection of the activations (see Fig.~\ref{fig:image04}a-l), we estimate that each $E(t)$ provides some $10^2$ patterns reminiscent of $g$, summed over all feature maps. On the other hand, for each $E(t)$ many patterns together are needed for an accurate prediction of $g$, which reduces the feature space from the number estimated before. To clarify this point we draw the analogy with having a picture of a cat, where a cat is defined by its eyes, ears, tail, fur, etc. Having only one of these features will not lead to an accurate prediction of the picture showing a cat. We therefore estimate that the total data size on which the CNN trains should be also of the order of $10^5$, which is on the low side of what is common for CNNs. 

The performance of the studied method is also determined by the experimental setup. The nebulized water does not always entirely end up in the area covered by the THz beam, thereby contributing to the measured gravimetric weight that is not seen by the THz beam. We expect that these errors are the cause of the horizontal lines in Fig.~\ref{fig:image02}d. Additionally, the accuracy of the gravimetric measurements of $g_b$ is rather low, mostly due to natural air convection and air streams caused by the nebulizers. However, the leaf cannot be placed in an enclosed box, as then the relative humidity gets spatial and temporal fluctuations with drastic consequences on $E(t)$. We estimate the absolute error on $g_b$ between 0.1 and 0.2\,mg.

Although these results indicate that our data-driven approaches provide a performant model to determine leaf wetness in a lab setting, the real baptism of fire for the methodology should be a test in the application environment. A leaf wetness sensor for the agriculture sector would need to continuously sense leaf wetness during the growth season on a representative size of the crop canopy. A reflection geometry would be preferred over a transmission one, and instead of a single leaf, many leaves will need to be probed simultaneously. In addition, leaves will not be clean and flat but occur with a large distribution of appearances. Moreover, the air can be highly humid or contain rain, and dust, and the beam path may be (temporally) disturbed by passing insects. We estimate that despite the reported performance of the algorithms presented in this work, this real setting, which may vary depending on the canopy, will be challenging for the described methods.

\section*{Conclusions}
We have studied the application of conventional machine learning methods to THz time-domain data to determine leaf wetness. Hereto, we experimentally acquired a large data set of 12,000 distinct time domain traces $E(t)$ corresponding to a distribution of water patterns on plant leaves. Using domain knowledge related to the light-matter interaction, we designed and trained a decision tree model and a convolutional neural network and gave special attention to feature extraction. Both models predict leaf wetness with an accuracy of about 4\,\%. The generalizability of the methods was evaluated on unseen datasets with increasing deviations from the training set. We observe a similarity between features important for CNN and those having a physical interpretation. In conclusion, conventional machine learning models can be of additional value when compared to model-based signal processing, especially in cases where the sample configuration is complex and well beyond a multilayer structure, although the variability with slightly different input data is an obstacle for practical application.

\section*{Methods}
\subsection*{Samples}
The plant material used in this work is \textit{Alliaria petiolata}, also known as \textit{garlic mustard}. The plant is widely spread in the Netherlands, in the wild, and in private gardens. The plant has been grown in the open ground, and leaves are harvested in May to June. Of the selected leaf, a circle of 30 mm diameter is cut and immediately embedded in between two plastic sheets each 0.08 mm thick to prevent it from fast drying out. The plastic sheets have not been sealed at their edges. During the measurement campaign, the leaf naturally dried out on the order of $18\%$ (that is, about 11 mg) in 5 days. This range of relative leaf water content corresponds to variations due to well-watered vs. severe-drought growth conditions.

To create moisture patterns on the plasticized leaf sample, two nebulizers (Medisana, Inhalator IN 500) are used. Two specially designed nozzles, which are symmetrically positioned perpendicular to the plane of incidence and facing the sample, ensure a homogeneous droplet pattern on the sample. The nebulizers are filled with distilled water.

\subsection*{Experimental setup}
The optical properties of a moistened plant leaf have been obtained by performing THz time-domain transmission spectroscopy (Toptica Teraflash Pro). The sample has been put out-of-focus such that the THz beam covers the surface area on which water is nebulized. The two nebulizers have been switched on at random to create a wide range of different droplet patterns. The gravimetric weight $g_b$ of the moist sample is recorded simultaneously with the THz data. Hereto, the plasticized leaf sample was mounted on a dedicated holder attached to a precision scale (Sartorius WZA224-L). In addition, an RGB picture is taken from each water pattern using an optical camera (Logitech Brio 4K). Simultaneously, also the absolute air humidity $a$ has been recorded. During the entire measurement campaign, which lasted 7 days, the temperature was 25.0~$^\circ$C $\pm$ 0.3~$^\circ$C, and the relative humidity was $42\% \pm 6\%$.

Training and inference of the described models are performed on a CPU (Intel i7-7700HQ @ 2.8 GHz), using 4/4 cores and multi-threading enabled.

\subsection*{Data acquisition}
Using the experimental setup described above, the transmitted electric field $E(t)$ has experimentally been determined of a leaf sample that has been moistened until a first droplet ran down the sample. A measurement sequence is divided into separate time slots: 0.5 s for nebulization, 1.0 s waiting time where the nebulization cloud deposits itself onto the sample and diffuses away from the beam path, and 1.5 s for acquiring 50 averages of $E(t)$, recording $g_b$ and $a$, and obtaining an RGB picture of the moisture pattern. For data set $\mathbf{E}(t)$ the THz beam shines onto the upper side of the leaf, whereas for data set $\mathbf{E'}(t)$ the THz beam shines onto the bottom side of the leaf. In both cases, the water pattern is directed towards the emitter side. $\mathbf{E}(t)$ contains a total of 10,609 different moisture patterns measured in 39 measurement series within 5 days. $\mathbf{E'}(t)$ contains 1,501 patterns recorded in 5 measurement series within 2 days. Each $E(t)$ contains 760 data points and spans a delay of 38 ps.

\subsection*{Data processing}
\label{sec:DataProcessing}
\subsubsection*{Decision trees}

Feature extraction is performed by fitting a polynomial function of order $n$ to a selected temporal range $m$ of $E(t)$. As $E(t)$ has a high temporal variation, and to keep $n$ low for optimal performance of the method, we choose several ranges to capture all relevant patterns of $E(t)$. The standard deviation $\sigma$ of $\mathbf{E}(t)$ is enhanced in the range (4, 22) ps (see Fig.~\ref{fig:image02}b) which we split into $m=4$ equal temporal regions (4, 7); (7, 10); (15.5, 18.5) and (18.5, 21.5) ps. As $E(t)$ for $g=0$, $E_0(t)$, manifests a slight temporal shift due to factors such as leaf water content, air humidity, and drift, the onset of all measured traces $E(t)$ occurs at a slightly different value of $t$. To correct for this, the onset of region $m=1$ is determined by the slope $d/dt$ of $E(t)/\max(E(t))$. Hereto, $E(t)$ has been interpolated to enhance the temporal precision. Determination of the order $n$ of each range has been done using a grid search algorithm, which is a brute-force search of a predefined set of options. Hereto, we calculate the L2 loss of the validation set after training on $n \in [0, 20]$. For the four different temporal regions of interest we obtain optimal polynomial fits with order $n_1 = 11, n_2 = 2, n_3 = 4$, and $n_4 = 8$, respectively. Subsequently, we eliminate recursive features from the input features using permutation importance as a ranking criterion. By shuffling a feature column in the input matrix, the correlation between the feature and target value is evaluated. As such, features with little correlation to the target value are iteratively eliminated, as well as features that are correlated to others that already show dependence on the target value. This reduces the number of input features and in turn, increases the predictive power of the validation set. In this way, the feature vectors obtained by applying the decision tree algorithm simultaneously to all regions $m$ read [$t^2$, $t^4$, $t_{\text{start}}$,~$a$], [bias,~$t^2$], [$t^1$,~$t^2$], [bias,~$t^2$,~$t^6$], for $m=1..4$ respectively. One can see that although both $a$ and $t_{\text{start}}$ are present within the features for $m=1$, they are eliminated from the features for $m>1$ due to mutual correlations.

To train a generalizable model, we use an ensemble of randomized decision trees, opposite to for instance a single tree which is prone to overfitting. The randomization is effectuated by the well-established principle of bootstrap aggregation (or bagging), where different subsets $D_i$ are sampled from training set $D$ with replacement, meaning that the same sample can be sampled multiple times. We further optimize the trees by tuning hyperparameters that control regularization, including the number of samples per tree $\textsc{n\_samples}$, the number of features per tree $\textsc{n\_features}$, and the maximum depth of branches of a tree $\textsc{max\_depth}$. 
After optimization of these hyperparameters using the mentioned grid search algorithm, we obtain $\textsc{n\_samples = 2000}$, 
and $\textsc{max\_features}$ turns out to be equal to $\textsc{n\_features}$ such that the decision tree becomes a bagged decision tree. The minimum number of samples after a split is set to 5 to smooth the predicted value of a tree. \cite{geurts_extremely_2006}

\subsection*{Convolutional Neural network}
The architecture of the CNN is shown in Fig.~\ref{fig:image04}m. The input data is inserted in a concatenation of six groups each containing a convolutional layer. This structure is motivated by a receptive field which increases with each deeper layer. Since deeper layers learn more complex patterns, the number of filters is increased by a factor of two for each layer. Moreover, the convolutional layers are (after batch normalization, see below) succeeded by a max-pooling layer that decreases the length by a factor of two. Given the kernel dimension $(1\times3)$ and that of the input vector $E(t)$ $(1\times760)$, we have chosen for zero padding the extremes of $E(t)$ with a single entry to have the dimension unaltered after convolution. After this feature extraction part, the output is flattened and feature $\mathbf{a}$ is appended before inserting it in a two-layer fully connected network. The entire network has 72,385 trainable parameters: 26,144 are responsible for feature detection and 46,241 for regression.

Input data needs to be normalized before feeding it into the network to ensure proper convergence during gradient descent, as steepest gradient descent algorithms do not possess the property of scale invariance. Normalization of the absolute humidity is performed as $a_i' = (a_i - \mu_{a})/\sigma_{a}$ where $\mu_{a}$ is the mean of $a$ with respect to the training population and $\sigma_{a}$ its standard deviation. For all $E(t)$, a uniform division factor of 4 is applied to ensure all data points are within the range $(-1, 1)$. For activation, we make use of the Rectified Linear Unit (ReLU) function, defined as $\varphi(h) = \max(0, h)$.

Convolutional networks require supervised learning to learn the intrinsic patterns to predict the target value. Supervised deep learning models use backpropagation to autonomously shape function $f$ to find relation $f:x \rightarrow y$, where $x$ is the input and $y$ the target. We have used the Adam optimizer to find the optimal model parameters. Adam is a simple and computationally efficient algorithm that combines the advantages of AdaGrad and RMSProp, resulting in an optimizer that is robust and well-suited to a wide range of non-convex optimization problems in the field of machine learning.\cite{kingma_adam_2017}

\bibliography{Agriculture.bib}

\begin{thebibliography}{10}
\urlstyle{rm}
\expandafter\ifx\csname url\endcsname\relax
  \def\url#1{\texttt{#1}}\fi
\expandafter\ifx\csname urlprefix\endcsname\relax\def\urlprefix{URL }\fi
\expandafter\ifx\csname doiprefix\endcsname\relax\def\doiprefix{DOI: }\fi
\providecommand{\bibinfo}[2]{#2}
\providecommand{\eprint}[2][]{\url{#2}}

\bibitem{koch_terahertz_2023}
\bibinfo{author}{Koch, M.}, \bibinfo{author}{Mittleman, D.~M.},
  \bibinfo{author}{Ornik, J.} \& \bibinfo{author}{Castro-Camus, E.}
\newblock \bibinfo{journal}{\bibinfo{title}{Terahertz time-domain
  spectroscopy}}.
\newblock {\emph{\JournalTitle{Nature Reviews Methods Primers}}}
  \textbf{\bibinfo{volume}{3}}, \bibinfo{pages}{48},
  \doiprefix\url{10.1038/s43586-023-00232-z} (\bibinfo{year}{2023}).

\bibitem{van_mechelen_industrial_2015}
\bibinfo{author}{van Mechelen, D.}
\newblock \bibinfo{journal}{\bibinfo{title}{An industrial {THz} killer
  application?}}
\newblock {\emph{\JournalTitle{Optics \& Photonics News}}}
  \textbf{\bibinfo{volume}{26}}, \bibinfo{pages}{16--18}
  (\bibinfo{year}{2015}).

\bibitem{leitenstorfer_2023_2023}
\bibinfo{author}{Leitenstorfer, A.} \emph{et~al.}
\newblock \bibinfo{journal}{\bibinfo{title}{The 2023 terahertz science and
  technology roadmap}}.
\newblock {\emph{\JournalTitle{Journal of Physics D: Applied Physics}}}
  \textbf{\bibinfo{volume}{56}}, \bibinfo{pages}{223001},
  \doiprefix\url{10.1088/1361-6463/acbe4c} (\bibinfo{year}{2023}).

\bibitem{van_mechelen_stratified_2014}
\bibinfo{author}{van Mechelen, J. L.~M.}, \bibinfo{author}{Kuzmenko, A.~B.} \&
  \bibinfo{author}{Merbold, H.}
\newblock \bibinfo{journal}{\bibinfo{title}{Stratified dispersive model for
  material characterization using terahertz time-domain spectroscopy}}.
\newblock {\emph{\JournalTitle{Optics Letters}}} \textbf{\bibinfo{volume}{39}},
  \bibinfo{pages}{3853--3856} (\bibinfo{year}{2014}).

\bibitem{s_recent_2022}
\bibinfo{author}{S, K.}, \bibinfo{author}{M, Y.}, \bibinfo{author}{Rawson, A.}
  \& \bibinfo{author}{C.~K, S.}
\newblock \bibinfo{journal}{\bibinfo{title}{Recent {Advances} in {Terahertz}
  {Time}-{Domain} {Spectroscopy} and {Imaging} {Techniques} for {Automation} in
  {Agriculture} and {Food} {Sector}}}.
\newblock {\emph{\JournalTitle{Food Analytical Methods}}}
  \textbf{\bibinfo{volume}{15}}, \bibinfo{pages}{498--526},
  \doiprefix\url{10.1007/s12161-021-02132-y} (\bibinfo{year}{2022}).

\bibitem{oerke_crop_2006}
\bibinfo{author}{Oerke, E.-C.}
\newblock \bibinfo{journal}{\bibinfo{title}{Crop losses to pests}}.
\newblock {\emph{\JournalTitle{The Journal of Agricultural Science}}}
  \textbf{\bibinfo{volume}{144}}, \bibinfo{pages}{31--43},
  \doiprefix\url{10.1017/S0021859605005708} (\bibinfo{year}{2006}).

\bibitem{bregaglio_multi_2011}
\bibinfo{author}{Bregaglio, S.}, \bibinfo{author}{Donatelli, M.},
  \bibinfo{author}{Confalonieri, R.}, \bibinfo{author}{Acutis, M.} \&
  \bibinfo{author}{Orlandini, S.}
\newblock \bibinfo{journal}{\bibinfo{title}{Multi metric evaluation of leaf
  wetness models for large-area application of plant disease models}}.
\newblock {\emph{\JournalTitle{Agricultural and Forest Meteorology}}}
  \textbf{\bibinfo{volume}{151}}, \bibinfo{pages}{1163--1172},
  \doiprefix\url{10.1016/j.agrformet.2011.04.003} (\bibinfo{year}{2011}).

\bibitem{huber_modeling_1992}
\bibinfo{author}{Huber, L.} \& \bibinfo{author}{Gillespie, T.~J.}
\newblock \bibinfo{journal}{\bibinfo{title}{Modeling {Leaf} {Wetness} in
  {Relation} to {Plant} {Disease} {Epidemiology}}}.
\newblock {\emph{\JournalTitle{Annual Review of Phytopathology}}}
  \textbf{\bibinfo{volume}{30}}, \bibinfo{pages}{553--577},
  \doiprefix\url{10.1146/annurev.py.30.090192.003005} (\bibinfo{year}{1992}).

\bibitem{goffart_potato_2022}
\bibinfo{author}{Goffart, J.-P.} \emph{et~al.}
\newblock \bibinfo{journal}{\bibinfo{title}{Potato production in northwestern
  europe (germany, france, the netherlands, united kingdom, belgium):
  Characteristics, issues, challenges and opportunities}}.
\newblock {\emph{\JournalTitle{Potato Research}}}
  \textbf{\bibinfo{volume}{65}}, \bibinfo{pages}{503--547},
  \doiprefix\url{10.1007/s11540-021-09535-8} (\bibinfo{year}{2022}).

\bibitem{li_non-invasive_2020}
\bibinfo{author}{Li, R.}, \bibinfo{author}{Lu, Y.}, \bibinfo{author}{Peters, J.
  M.~R.}, \bibinfo{author}{Choat, B.} \& \bibinfo{author}{Lee, A.~J.}
\newblock \bibinfo{journal}{\bibinfo{title}{Non-invasive measurement of leaf
  water content and pressure–volume curves using terahertz radiation}}.
\newblock {\emph{\JournalTitle{Scientific Reports}}}
  \textbf{\bibinfo{volume}{10}}, \bibinfo{pages}{21028},
  \doiprefix\url{10.1038/s41598-020-78154-z} (\bibinfo{year}{2020}).

\bibitem{gente_determination_2013}
\bibinfo{author}{Gente, R.} \emph{et~al.}
\newblock \bibinfo{journal}{\bibinfo{title}{Determination of {Leaf} {Water}
  {Content} from {Terahertz} {Time}-{Domain} {Spectroscopic} {Data}}}.
\newblock {\emph{\JournalTitle{Journal of Infrared, Millimeter, and Terahertz
  Waves}}} \textbf{\bibinfo{volume}{34}}, \bibinfo{pages}{316--323},
  \doiprefix\url{10.1007/s10762-013-9972-8} (\bibinfo{year}{2013}).

\bibitem{gente_monitoring_2015}
\bibinfo{author}{Gente, R.} \& \bibinfo{author}{Koch, M.}
\newblock \bibinfo{journal}{\bibinfo{title}{Monitoring leaf water content with
  {THz} and sub-{THz} waves}}.
\newblock {\emph{\JournalTitle{Plant Methods}}} \textbf{\bibinfo{volume}{11}},
  \bibinfo{pages}{15}, \doiprefix\url{10.1186/s13007-015-0057-7}
  (\bibinfo{year}{2015}).

\bibitem{singh_three-dimensional_2020}
\bibinfo{author}{Singh, A.~K.}, \bibinfo{author}{Pérez-López, A.~V.},
  \bibinfo{author}{Simpson, J.} \& \bibinfo{author}{Castro-Camus, E.}
\newblock \bibinfo{journal}{\bibinfo{title}{Three-dimensional water mapping of
  succulent {Agave} victoriae-reginae leaves by terahertz imaging}}.
\newblock {\emph{\JournalTitle{Scientific Reports}}}
  \textbf{\bibinfo{volume}{10}}, \bibinfo{pages}{1404},
  \doiprefix\url{10.1038/s41598-020-58277-z} (\bibinfo{year}{2020}).

\bibitem{rowlandson_reconsidering_2015}
\bibinfo{author}{Rowlandson, T.} \emph{et~al.}
\newblock \bibinfo{journal}{\bibinfo{title}{Reconsidering {Leaf} {Wetness}
  {Duration} {Determination} for {Plant} {Disease} {Management}}}.
\newblock {\emph{\JournalTitle{Plant Disease}}} \textbf{\bibinfo{volume}{99}},
  \bibinfo{pages}{310--319}, \doiprefix\url{10.1094/PDIS-05-14-0529-FE}
  (\bibinfo{year}{2015}).
\newblock \bibinfo{note}{Publisher: Scientific Societies}.

\bibitem{van_mechelen_thickness_2021}
\bibinfo{author}{van Mechelen, J. L.~M.}, \bibinfo{author}{Frank, A.} \&
  \bibinfo{author}{Maas, D. J. H.~C.}
\newblock \bibinfo{journal}{\bibinfo{title}{Thickness sensor for drying paints
  using {THz} spectroscopy}}.
\newblock {\emph{\JournalTitle{Optics Express}}} \textbf{\bibinfo{volume}{29}},
  \bibinfo{pages}{7514}, \doiprefix\url{10.1364/OE.418809}
  (\bibinfo{year}{2021}).

\bibitem{park_machine_2021}
\bibinfo{author}{Park, H.} \& \bibinfo{author}{Son, J.-H.}
\newblock \bibinfo{journal}{\bibinfo{title}{Machine {Learning} {Techniques} for
  {THz} {Imaging} and {Time}-{Domain} {Spectroscopy}}}.
\newblock {\emph{\JournalTitle{Sensors}}} \textbf{\bibinfo{volume}{21}},
  \bibinfo{pages}{1186}, \doiprefix\url{10.3390/s21041186}
  (\bibinfo{year}{2021}).

\bibitem{wang_terahertz_2020}
\bibinfo{author}{Wang, Y.} \emph{et~al.}
\newblock \bibinfo{journal}{\bibinfo{title}{Terahertz spectroscopic diagnosis
  of early blast-induced traumatic brain injury in rats}}.
\newblock {\emph{\JournalTitle{Biomedical Optics Express}}}
  \textbf{\bibinfo{volume}{11}}, \bibinfo{pages}{4085},
  \doiprefix\url{10.1364/BOE.395432} (\bibinfo{year}{2020}).

\bibitem{cao_terahertz_2020}
\bibinfo{author}{Cao, C.}, \bibinfo{author}{Zhang, Z.}, \bibinfo{author}{Zhao,
  X.} \& \bibinfo{author}{Zhang, T.}
\newblock \bibinfo{journal}{\bibinfo{title}{Terahertz spectroscopy and machine
  learning algorithm for non-destructive evaluation of protein conformation}}.
\newblock {\emph{\JournalTitle{Optical and Quantum Electronics}}}
  \textbf{\bibinfo{volume}{52}}, \bibinfo{pages}{225},
  \doiprefix\url{10.1007/s11082-020-02345-1} (\bibinfo{year}{2020}).

\bibitem{wang_automatic_2021}
\bibinfo{author}{Wang, Q.} \emph{et~al.}
\newblock \bibinfo{journal}{\bibinfo{title}{Automatic defect prediction in
  glass fiber reinforced polymer based on {THz}-{TDS} signal analysis with
  neural networks}}.
\newblock {\emph{\JournalTitle{Infrared Physics \& Technology}}}
  \textbf{\bibinfo{volume}{115}}, \bibinfo{pages}{103673},
  \doiprefix\url{10.1016/j.infrared.2021.103673} (\bibinfo{year}{2021}).

\bibitem{li_nondestructive_2022}
\bibinfo{author}{Li, R.} \emph{et~al.}
\newblock \bibinfo{journal}{\bibinfo{title}{Nondestructive {Evaluation} of
  {Thermal} {Barrier} {Coatings} {Thickness} {Using} {Terahertz}
  {Time}-{Domain} {Spectroscopy} {Combined} with {Hybrid} {Machine} {Learning}
  {Approaches}}}.
\newblock {\emph{\JournalTitle{MDPI}}} \textbf{\bibinfo{volume}{12}},
  \bibinfo{pages}{1875}, \doiprefix\url{10.3390/coatings12121875}
  (\bibinfo{year}{2022}).

\bibitem{mao_convolutional_2020}
\bibinfo{author}{Mao, Q.} \emph{et~al.}
\newblock \bibinfo{journal}{\bibinfo{title}{Convolutional neural network model
  based on terahertz imaging for integrated circuit defect detections}}.
\newblock {\emph{\JournalTitle{Optics Express}}} \textbf{\bibinfo{volume}{28}},
  \bibinfo{pages}{5000}, \doiprefix\url{10.1364/OE.384146}
  (\bibinfo{year}{2020}).

\bibitem{wang_convolutional_2021}
\bibinfo{author}{Wang, C.} \emph{et~al.}
\newblock \bibinfo{journal}{\bibinfo{title}{Convolutional {Neural}
  {Network}-{Based} {Terahertz} {Spectral} {Classification} of {Liquid}
  {Contraband} for {Security} {Inspection}}}.
\newblock {\emph{\JournalTitle{IEEE Sensors Journal}}}
  \textbf{\bibinfo{volume}{21}}, \bibinfo{pages}{18955--18963},
  \doiprefix\url{10.1109/JSEN.2021.3086478} (\bibinfo{year}{2021}).

\bibitem{koumans_sensing_2022}
\bibinfo{author}{Koumans, M.}, \bibinfo{author}{Perez-Casanova, A.} \&
  \bibinfo{author}{Van~Mechelen, J. L.~M.}
\newblock \bibinfo{title}{Sensing moisture patterns using terahertz
  spectroscopy}.
\newblock In \emph{\bibinfo{booktitle}{2022 47th {International} {Conference}
  on {Infrared}, {Millimeter} and {Terahertz} {Waves} ({IRMMW}-{THz})}},
  \bibinfo{pages}{1--2}, \doiprefix\url{10.1109/IRMMW-THz50927.2022.9895781}
  (\bibinfo{publisher}{IEEE}, \bibinfo{address}{Delft, Netherlands},
  \bibinfo{year}{2022}).

\bibitem{ioffe_batch_2015}
\bibinfo{author}{Ioffe, S.} \& \bibinfo{author}{Szegedy, C.}
\newblock \bibinfo{title}{Batch {Normalization}: {Accelerating} {Deep}
  {Network} {Training} by {Reducing} {Internal} {Covariate} {Shift}}
  (\bibinfo{year}{2015}).
\newblock \bibinfo{note}{ArXiv:1502.03167 [cs]}.

\bibitem{Jordens2009}
\bibinfo{author}{Jördens, C.}, \bibinfo{author}{Scheller, M.},
  \bibinfo{author}{Breitenstein, B.}, \bibinfo{author}{Selmar, D.} \&
  \bibinfo{author}{Koch, M.}
\newblock \bibinfo{journal}{\bibinfo{title}{Evaluation of leaf water status by
  means of permittivity at terahertz frequencies}}.
\newblock {\emph{\JournalTitle{Journal of Biological Physics}}}
  \textbf{\bibinfo{volume}{35}}, \bibinfo{pages}{255--264},
  \doiprefix\url{10.1007/s10867-009-9161-0} (\bibinfo{year}{2009}).

\bibitem{geurts_extremely_2006}
\bibinfo{author}{Geurts, P.}, \bibinfo{author}{Ernst, D.} \&
  \bibinfo{author}{Wehenkel, L.}
\newblock \bibinfo{journal}{\bibinfo{title}{Extremely randomized trees}}.
\newblock {\emph{\JournalTitle{Machine Learning}}}
  \textbf{\bibinfo{volume}{63}}, \bibinfo{pages}{3--42},
  \doiprefix\url{10.1007/s10994-006-6226-1} (\bibinfo{year}{2006}).

\bibitem{kingma_adam_2017}
\bibinfo{author}{Kingma, D.~P.} \& \bibinfo{author}{Ba, J.}
\newblock \bibinfo{title}{Adam: {A} {Method} for {Stochastic} {Optimization}}
  (\bibinfo{year}{2017}).
\newblock \bibinfo{note}{ArXiv:1412.6980 [cs]}.

\end{thebibliography}

%\noindent LaTeX formats citations and references automatically using the bibliography records in your .bib file, which you can edit via the project menu. Use the cite command for an inline citation, e.g.  \cite{Hao:gidmaps:2014}.

%For data citations of datasets uploaded to e.g. \emph{figshare}, please use the \verb|howpublished| option in the bib entry to specify the platform and the link, as in the \verb|Hao:gidmaps:2014| example in the sample bibliography file.

\section*{Acknowledgements}

D.v.M. and B.D. are grateful to fruitful discussions with Niels Anten (Wageningen University, the Netherlands), and for the facilitating role of research program Synergia - SYstem change for New Ecology-based and Resource efficient Growth with high tech In Agriculture, financed by NWO, the Dutch Research Council (project number 17626), industrial and scientific/research partners, as well as to Rik Vullings (Eindhoven University of Technology) for proofreading the manuscript.

\section*{Author contributions statement}

D.v.M. designed and supervised the project, B.D. conceived the problem statement, M.K., D.M, and H.M made the experimental setup and conducted the experiments, M.K. analysed the results. D.v.M, M.K. and B.D. wrote the manuscript.  All authors reviewed the manuscript. 

\section*{Additional information}
The authors declare no competing interests.

%To include, in this order: \textbf{Accession codes} (where applicable); \textbf{Competing interests} (mandatory statement). 

%The corresponding author is responsible for submitting a \href{http://www.nature.com/srep/policies/index.html#competing}{competing interests statement} on behalf of all authors of the paper. This statement must be included in the submitted article file.

%\begin{figure}[ht]
%\centering
%\includegraphics[width=\linewidth]{stream}
%\caption{Legend (350 words max). Example legend text.}
%\label{fig:stream}
%\end{figure}

%\begin{table}[ht]
%\centering
%\begin{tabular}{|l|l|l|}
%\hline
%Condition & n & p \\
%\hline
%A & 5 & 0.1 \\
%\hline
%B & 10 & 0.01 \\
%\hline
%\end{tabular}
%\caption{\label{tab:example}Legend (350 words max). Example legend text.}
%\end{table}

%Figures and tables can be referenced in LaTeX using the ref command, e.g. Figure \ref{fig:stream} and Table \ref{tab:example}.

\end{document}